\documentclass[acmlarge]{acmart}

\newcommand{\eg}{\textit{e}.\textit{g}.}
\newcommand{\ie}{\textit{i}.\textit{e}.}
\newcommand{\etal}{\textit{et al}.}
\newcommand{\parafold}{{\small ParaFold}}
\AtBeginDocument{%
  \providecommand\BibTeX{{%
    \normalfont B\kern-0.5em{\scshape i\kern-0.25em b}\kern-0.8em\TeX}}}

\begin{document}
\title{{\tt ParaFold}: Paralleling AlphaFold for Large-Scale Predictions}

\author{Bozitao Zhong}
\author{Xiaoming Su}
\author{Minhua Wen}
\author{Sicheng Zuo}
\affiliation{%
  \institution{Center for High Performance Computing, Shanghai Jiao Tong University}
  \city{Shanghai}
  \country{China}}
\email{james@sjtu.edu.cn}  

\author{Liang Hong}
\affiliation{%
  \institution{Institute of Natural Sciences, Shanghai Jiao Tong University}
  \city{Shanghai}
  \country{China}}

\author{James Lin}
\authornote{Corresponding Author}
\affiliation{%
  \institution{Center for High Performance Computing, Shanghai Jiao Tong University}
  \city{Shanghai}
  \country{China}}
\email{james@sjtu.edu.cn}  


\begin{abstract}
  AlphaFold developed by DeepMind predicts protein structures from the amino acid sequence at or near experimental resolution, solving the 50-year-old protein folding challenge, leading to progress by transforming large-scale genomics data into protein structures. AlphaFold will also greatly change the scientific research model from low-throughput to high-throughput manner. The AlphaFold framework is a mixture of two types of workloads: 1) MSA construction based on CPUs and 2) model inference on GPUs. The first CPU stage dominates the overall runtime, taking up to hours for a single protein due to the large database sizes and I/O bottlenecks. However, GPUs in this CPU stage remain idle, resulting in low GPU utilization and restricting the capacity of large-scale structure predictions. Therefore, we proposed {\small ``ParaFold"}, an open-source parallel version of AlphaFold for high throughput protein structure predictions. \parafold{} separates the CPU and GPU parts to enable large-scale structure predictions and to improve GPU utilization. \parafold{} also effectively reduces the CPU and GPU runtime with two optimizations without compromising the quality of prediction results: using multi-threaded parallelism on CPUs and using optimized JAX compilation on GPUs. We evaluated \parafold{} with three datasets of different size and protein lengths. With the small dataset, we evaluated the accuracy and efficiency of optimizations on CPUs and GPUs; With the medium dataset, we demonstrated a typical usage of structure predictions of proteins of different sizes ranging from 77 to 734 residues; With the large dataset, we showed the large-scale prediction capability by running model 1 inferences of $\sim$20,000 small proteins in five hours on one NVIDIA DGX-2. Using the JAX compile optimization, ParaFold attained a 13.8X average speedup over AlphaFold. \parafold{} offers a rapid and effective approach for high-throughput structure predictions, leveraging the predictive power by running on supercomputers, with shorter time, and at a lower cost. The development of \parafold{} will greatly speed up high-throughput studies and render the protein “structure-omics” feasible.
\end{abstract}



\maketitle

\section{Introduction}


Accurate determination of the 3D atomic structure of biomolecules is of crucial importance for various biomedical and bioengineering applications including protein design, drug design, diagnosing of diseases, \textit{etc.} In the past, this task was mainly achieved using expensive experimental methods, such as X-ray crystallography, Cryo-EM, NMR, \textit{etc.}, in a low-throughput manner. The recent success of AlphaFold \footnote{AlphaFold v2.0. For expediency, we refer to this model simply as AlphaFold throughout the rest of this paper.} \cite{jumper2021highly,alphafoldgithub,Evans2021.10.04.463034} greatly changes the paradigm and catalyzes the transition from the experiment-dominated solution to the AI driven process. Not only does AlphaFold render many research groups and industrial players without access to the expensive experimental tools the capability to explore the structure of biomolecules they concerned, it will also greatly change the scientific research model from low-throughput to high-throughput manner \cite{tunyasuvunakool2021highly}. For example, by using AlphaFold, one can compare the structures of thousands of pairs of proteins of similar functions between two different bacterial or explore how the structure of one single protein varies among many thousands of species along the evolution tree. The availability of large numbers of predicted protein structures provides a veritable cornucopia of data to be exploited, analysed and mined by structural bioinformaticians. The breakthrough will lead to an encyclopedia of the structures of all known protein domains, enabling a complete structural coverage of proteomes. AlphaFold is most likely to be the start of a revolution based on data-driven prediction in biology and medicine \cite{thornton2021alphafold}.


The overall prediction process of AlphaFold consists of two main stages: MSA (multiple sequence alignments) construction and model inference. 1) For the MSA construction stage, AlphaFold uses the input sequence and queries databases to generate an MSA and a list of templates. 2) For the model inference stage, AlphaFold extracts the information from the MSA using a new Evoformer architecture, and passes that information to the structure module. The structure module takes the representation and builds a 3D structure model followed by local refinement to provide the final prediction. 


AlphaFold operates the end-to-end prediction in a single task, yet the two main stages require different resources. In the first stage, MSA construction runs on CPUs only, while in the second stage, model inference performs best on GPUs. Therefore, for speed and convenience, AlphaFold runs prediction tasks on GPUs. Meanwhile, the first CPU stage dominates the overall runtime. Due to the large database sizes and I/O bottlenecks, MSA construction can take up to hours for a single protein \cite{jumper2021highly,tunyasuvunakool2021highly,mirdita2021colabfold}.

However, GPUs remain idle in the MSA construction stage, accounting for a large part of the total runtime. AlphaFold was mainly designed to predict single protein target in CASP14 (an independently assessed biennial community-wide competition) \cite{pereira2021high}. In order to rapidly explore large number of protein molecules like proteome or design protein libraries, a parallel optimized version of AlphaFold for high throughput use is highly desired.


Therefore, we proposed \parafold{}, a parallel version of AlphaFold with separated and optimized CPU and GPU tasks. Our work consists of two parts: pipeline design and performance optimization. First, we optimized the pipeline by segregating the CPU and GPU workload into individual jobs with proper resources. Second, we applied two performance optimizations to the pipeline: 1) Parallel acceleration on CPUs with three MSA searches running in parallel. 2) JAX \cite{bradbury2018jax} compile optimization on GPUs by avoiding recompilation in batch inferences.


We evaluated \parafold{} with three datasets (small, medium, and large). 1) The large dataset consists of $\sim$20,000 small proteins (50 residues in length). Structure predictions were performed on one NVIDIA DGX-2 multi-GPU system (16 V100/32G GPUs) and 10,400 cores of Intel Xeon Gold 6248 CPU on a cluster. The result showed that, \parafold{} took only five hours to complete model 1 inference of $\sim$20,000 proteins on one NVIDIA DGX-2. The GPU runtime for these proteins was only 1/241 of the total GPU time needed by AlphaFold. 2) The medium dataset consists of 100 proteins of various lengths to represent a typical use of protein predictions and to illustrate the speed of \parafold{}. 3) The small dataset contains four proteins. We evaluated the accuracy and efficiency of optimizations in  \parafold{} by comparing to those of AlphaFold with this small dataset.

\parafold{} is an open-sourced project, and the code is available at GitHub \cite{zurichogithub,hpcgithub}. As far as we know, \parafold{} is the first open source parallel version of AlphaFold for large-scale predictions. Although DeepMind used a proteome-scale pipeline to predict structures of the UniProt human reference proteome, the script and details of the pipeline was not published  \cite{tunyasuvunakool2021highly}. 


Concisely, this work makes the following contributions:

\begin{itemize}
\item We proposed the first pipeline optimized for rapid and large-scale protein structure predictions based on AlphaFold, suitable for running on supercomputers.
\item We effectively reduced the runtime on CPUs using multi-threaded parallelism and on GPUs using optimized JAX compilation, without compromising the quality of results.
\item We reported the first accomplishment of model 1 inferences of $\sim$20,000 small proteins in five hours on one NVIDIA DGX-2.
\end{itemize}


\section{Background}

Proteins are essential to life, and understanding their structure is a key step towards understanding and modifying their function. For decades, researchers deciphered protein 3D structures using experimental techniques such as X-ray crystallography or cryo-electron microscopy (cryo-EM). But such methods can take months or years of experimental process. Structures have been solved for only about 170,000 of the more than 200 million proteins discovered across life forms \cite{burley2021rcsb,thornton2021alphafold}. It has been considered the holy grail of biology to predicting 3D structures of proteins based solely on amino acid sequences \cite{jumper2021applying}. Many computational approaches have been developed, focusing on either thermodynamics or evolutionary approaches \cite{kuhlman2019advances}. However, all of them failed to live up to expectations until recently AlphaFold was entered into the CASP14 assessment. AlphaFold achieved a median score of 92.4 GDT-TS overall across all targets, with an average error of 1.6 Angstroms — about the radius of an atom \cite{jumper2021applying}. The AlphaFold models will be used in exactly the same way as experimental structural data (and indeed will be used to help determine low-resolution experimental structures) \cite{akdel2021structural}.

Briefly, the operation of AlphaFold falls into two parts (Fig. 1): 1) MSA construction on CPUs and 2) model inference on GPUs.

\begin{figure}[h]
  \centering
  \includegraphics[width=0.9\textwidth]{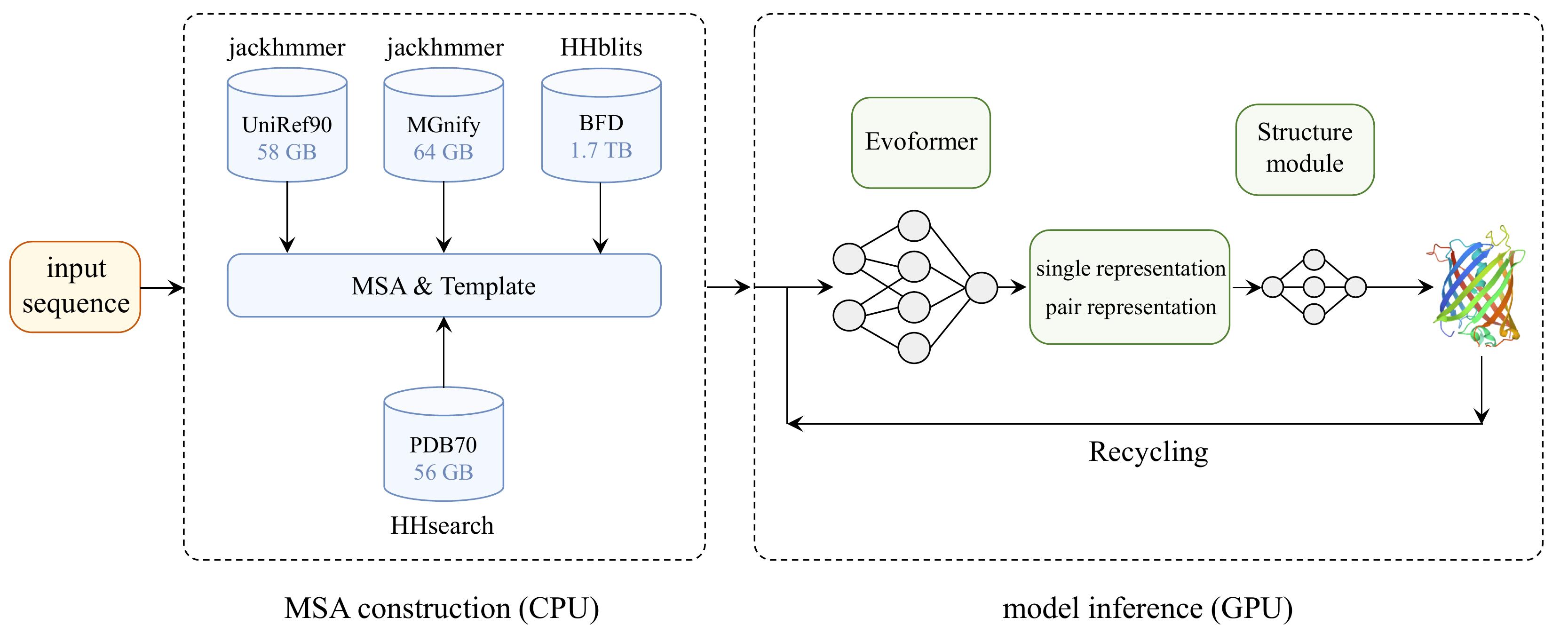}
  \caption{The AlphaFold structure prediction process consists of two main steps: 1) MSA construction using CPUs and 2) Model inference using GPUs.}
  \label{fig:alphafoldpipeline}
\end{figure}

First, for the CPU part, AlphaFold uses the input amino acid sequence to search through several protein sequence databases, and constructs an MSA for query sequence. AlphaFold also tries to identify proteins that may have a similar structure to the input (“templates”), and constructs an initial representation of the structure, which it calls the “pair representation”. To build diverse MSAs, large collections of protein sequences from public reference and environmental databases \cite{uniprot2019uniprot,steinegger2018clustering,steinegger2019protein,mitchell2020mgnify} are searched by AlphaFold using the sensitive homology detection methods jackhmmer \cite{eddy2011accelerated} and HHblits \cite{steinegger2019protein}. Specifically, AlphaFold uses jackhmmer for MSA search on Uniref90 \cite{suzek2015uniref} and clustered MGnify \cite{mitchell2020mgnify}, and uses HHBlits for MSA search on BFD  \cite{jumper2021applying} + Uniclust30 \cite{mirdita2017uniclust}, and HHSearch \cite{steinegger2019protein} for template search against PDB70 \cite{zimmermann2018completely}. AlphaFold restricts itself to 8 CPU cores for jackhmmer and 4 CPU cores for HHblits to process one query. Due to the large database sizes (over 2 TB) and the high number of random file accesses, the MSA search can take up to hours for a single prediction \cite{jumper2021highly,tunyasuvunakool2021highly,mirdita2021colabfold}. 

Second, for the GPU part, AlphaFold takes the features generate from MSA and the templates, and passes them through a complicated neural network. The objective of this neural network is to refine the representations for both the MSA and the pair interactions, but also to iteratively exchange information between them. Then it moves to the second neural network that produces a structure. It takes the refined “MSA representation” and “pair representation”, and leverages them to construct a 3D model of the structure. After generating a final structure, it will take all the information and pass it back to the beginning of the Evoformer blocks, in a “recycling” procedure to further refine the structure predictions. The model is trained end-to-end with gradients propagating from the predicted structure through the entire network.

AlphaFold provides 5 models which were used during CASP14 and were extensively validated for structure prediction quality, as well as 5 pTM models \cite{jumper2021highly}, which were fine-tuned to produce pTM (predicted TM-score) and predicted aligned error values alongside their structure predictions.


\section{Research Motivation}

\begin{figure}[h]
  \centering
  \includegraphics[width=0.7\textwidth]{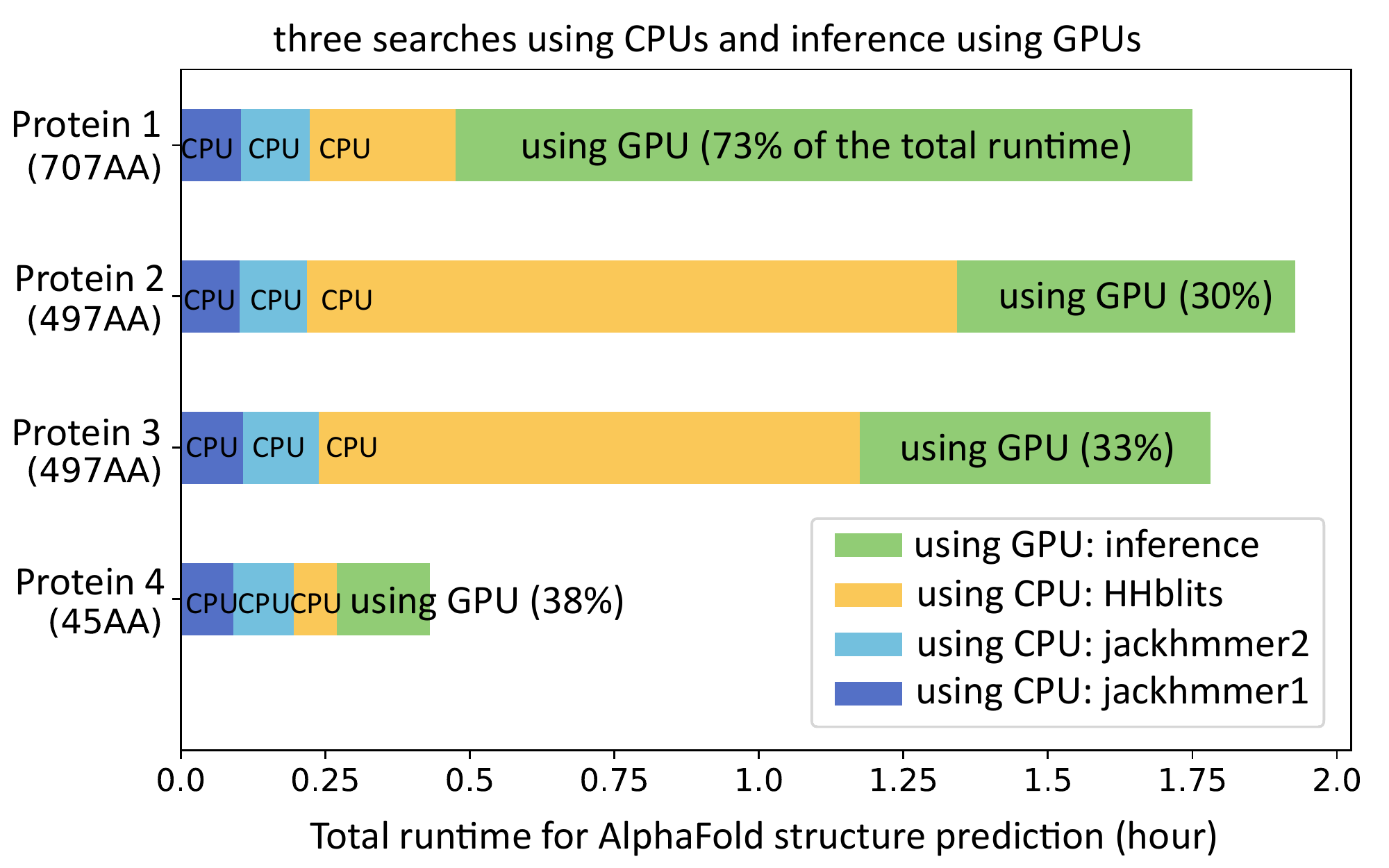}
  \caption{AlphaFold runtime for four proteins on one V100 GPU. The predictions were run entirely on the V100 GPU. However, the first three MSA searches (jackhmmer1, jackhmmer2, and HHblits) in the MSA construction stage used only CPUs, and GPUs remained idle in this stage. The green-colored inference stage was the only procedure that needed GPUs.}
  \label{fig:profiling}
\end{figure}

To explore the potential for optimizing for large-scale protein predictions, and for improving the GPU utilization,  we ran structure prediction of four proteins of lengths ranging from 45 to 707 residues on one V100/32G GPU (Fig \ref{fig:profiling}). Because the current estimate for the average protein length of all proteins is around 300 residues ($\eg$, the median length of the proteins annotated among Eukaryotes is 361 residues, Bacteria 267 residues, and Archaea 247 residues) \cite{brocchieri2005protein}, the four proteins with an average length of 436 residues in our test represented a typical usage in AlphaFold.

Fig \ref{fig:profiling} shows that, the predictions were run entirely on one V100/32G GPU, with runtime up to hours depending on the length of the proteins. The V100 GPU remained idle during the first three MSA searches, because these MSA searches used only CPUs. It was the fourth procedure (inference) that really needed GPUs. The real GPU workload accounted for 73\% of the total runtime for protein 1, 30\% for protein 2, 33\% for protein 3, and 38\% for protein 4. It means that the GPU utilization for these four proteins ranged from 33\% to 73\%.

To avoid low GPU utilization in AlphaFold, the three MSA searches in the CPU stage could be separated from the whole GPU workload. For high throughout predictions, these three CPU operations in AlphaFold could be scheduled to run on multiple CPU nodes on supercomputers. Furthermore, the three CPU operations could be arranged in parallel to reduce the CPU runtime.


\section{Design of ParaFold}

Our work on \parafold{} consists of two parts: pipeline design and performance optimization. First, we designed the pipeline for large-scale structure predictions. Second, we applied two performance optimizations on the pipeline to speedup the CPU and GPU operations.

\subsection{Pipeline for large-scale structure predictions}

By segregating the CPU and GPU workload into individual jobs with proper resources, we developed an efficient and scalable pipeline in \parafold{} for high throughput use. \parafold{} first runs the MSA construction on CPU nodes, then executes the model inference on GPUs. \parafold{} allows us to run large-scale protein predictions on supercomputer, with shorter time, and at a lower cost.

\begin{figure}[h]
  \centering
  \includegraphics[width=0.9\textwidth]{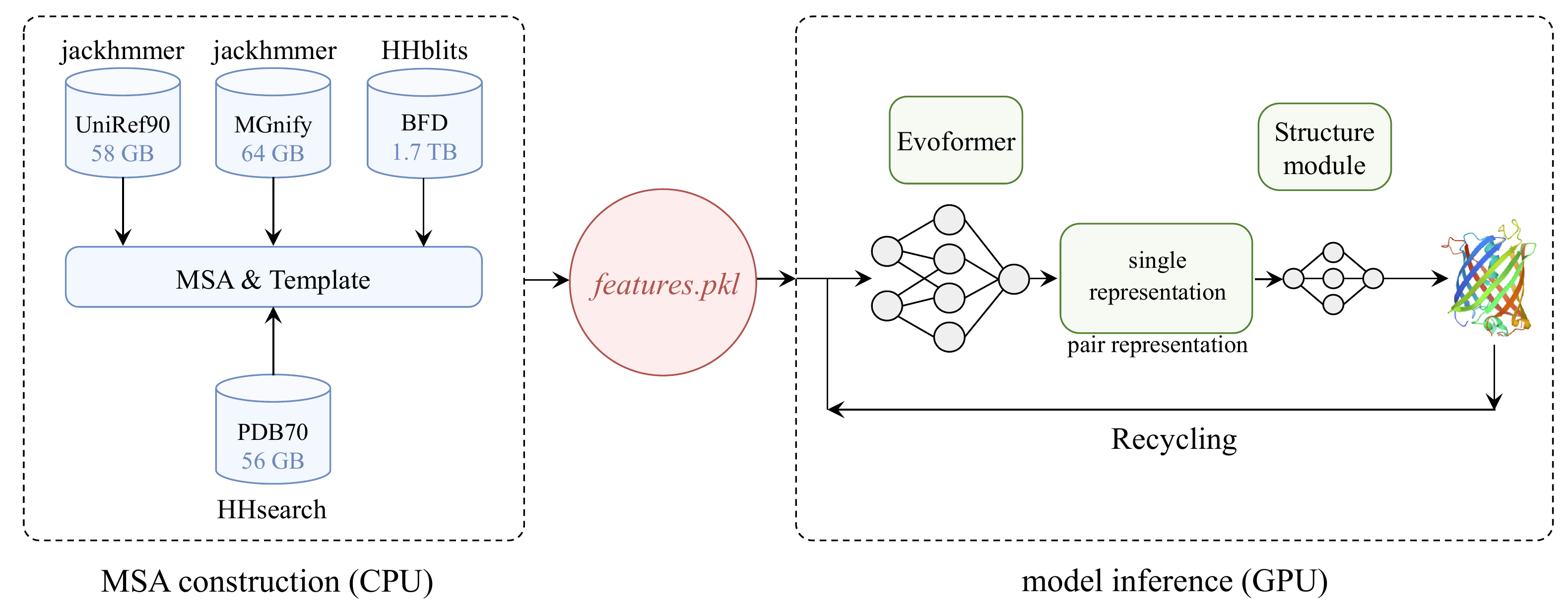}
  \caption{\parafold{} works by checking the existence of file $\textit features.pkl$. \parafold{} distributes the first stage jobs to CPUs. Once the file $\textit features.pkl$ is generated, the second stage of model inference on GPUs starts. \parafold{} also supports to run entirely on GPUs if the prediction job is submitted to GPUs but without the existence of $\textit features.pkl$.}
  \label{fig:split}
\end{figure}

As shown in Fig \ref{fig:split}, \parafold{} works by checking whether a file named $\textit features.pkl$ exists. $\textit features.pkl$ stores the MSA and structure template search results obtained on the CPUs and passes them to the neural network prediction on GPUs, and serves as the connection between the CPU and GPU stage in the whole end-to-end neural network process.

\parafold{} distributes the first stage jobs to CPUs. These CPU jobs usually take a few minutes or hours to complete. Once the file $\textit features.pkl$ is generated by CPUs, the second stage of model inference on GPUs starts. \parafold{} also supports to run entirely on GPUs, like AlphaFold, if the prediction job is submitted to GPUs but without the existence of $\textit features.pkl$.

\subsection{Performance optimizations}

We applied two performance optimizations on \parafold{}: one to run multiple MSA searches in parallel for speedup on CPUs, and the other to avoid JAX recompilation for speedup on GPUs.

\subsubsection{CPU acceleration}

\begin{figure}[h]
  \centering
  \includegraphics[width=1\textwidth]{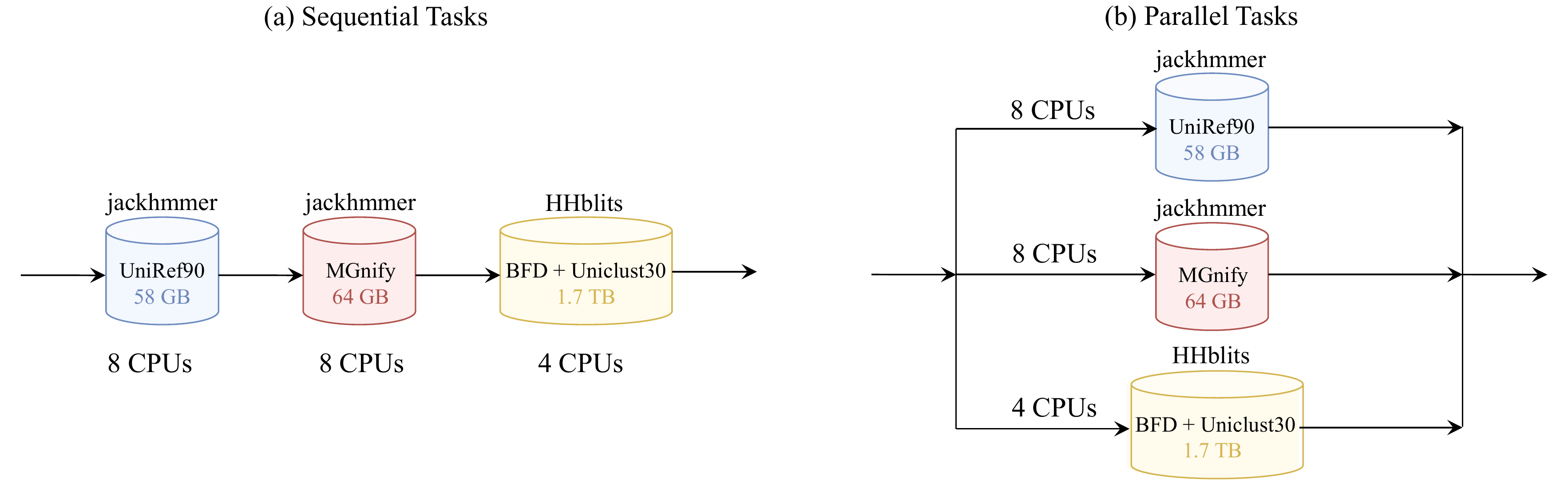}
  \caption{Acceleration of MSA searches procedure on CPUs by running in parallel.}
  \label{fig:search}
\end{figure}

To accelerate the CPU stage, three independent sequential MSA searches can be arranged in parallel (Fig \ref{fig:search}). Due to the limited CPU cores accompanying GPUs, AlphaFold restricts itself to 8 CPU cores for jackhmmer and 4 CPU cores for HHblits to process one query. With unlimited processors on CPU nodes, \parafold{} enhances the speed of MSA construction by orchestrating these searches in parallel using a total of 20 cores, $\ie$ 8 CPUs for jackhmmer1, 8 CPUs for jackhmmer2, and 4 CPUs for HHblits. 

In AlphaFold, as shown in Fig \ref{fig:search}(a), three datasets UniRef90, MGnify, and BFD sequentially were searched by jackhmmer1, jackhmmer2, and HHblits, respectively. In contrast, In \parafold{}, we simultaneously started three processes through Python's multiprocessing library to perform MSA searches in parallel, as shown in Fig \ref{fig:search}(b).

\subsubsection{GPU acceleration}

To accelerate the GPU stage, \parafold{} provides optimized batch inference script to avoid JAX recompilation for proteins of similar length.

In AlphaFold, JAX compiled the neural network to be specialized to exactly the size of the protein, MSA, and templates. For a single protein, the compile time was 5 when computing 5 models (model 1-5), and 10 when computing 10 models (model 1-5, pTM model 1-5). For large proteins, the compile time is a negligible fraction of the runtime, but it may become more significant for small proteins \cite{alphafoldgithub}.

In \parafold{}, we optimized JAX to avoid recompilation in batch inferences of proteins of similar length. All the proteins of similar length were sorted by length, and JAX compiled for the first and largest protein only, and all the other proteins shared the model compiled, without triggering recompilation. Therefore, in batch inferences of $N$ proteins, the compile time was thus reduced from 5$N$ to 5 when computing 5 models, and 10$N$ to 10 when computing 10 models.


\section{Evaluation}

\subsection{Experimental Setup}

\subsubsection{Hardware and Software}

We used one NVIDIA DGX-2 multi-GPU system (16 V100/32G GPUs), and 10,400 cores of Intel Xeon Gold 6248 2.5 GHz CPUs with 192 GB RAM on the $\pi$ 2.0 supercomputer in Shanghai Jiao Tong University. 

The version of AlphaFold is v2.0.1, released in Oct. 2021. The \parafold{} is v1.0. Both of AlphaFold and \parafold{} used CASP14 preset and default databases stored on the Lustre file system. Templates and Amber relaxation were not used in the prediction.

\subsubsection{Test Cases}

Three protein datasets of small, medium, and large size were used to evaluate the efficiency and operating performance of \parafold{}, as listed in Table\ref{tab:bench}. 

\begin{itemize}
  \item The small dataset consists of four proteins, the mutants GA98, GB98, 2LHE, and 2LHG. These four proteins differing in single mutation positions, with a chain length of 56 amino acids, represent diverse 3D structures: monomeric 3$\alpha$ and  4$\beta$ + $\alpha$ folds  \cite{alexander2009minimal,he2012mutational}.
  \item The medium dataset consists of 100 proteins  of varying lengths ranging from 77 to 734 residues, with an average of 296 residues. It is a randomly selected subset of sequences from a archaea proteome \cite{zhao2015thermococcus}. 
  \item The large dataset contains 19,704 small proteins from the a \textit{de novo} designed dataset \cite{rocklin2017global}. These \textit{de novo} proteins are of the same length of 50 residues.
\end{itemize}

\begin{table}
\caption{The benchmark datasets used for ParaFold evaluation.}
\label{tab:bench}
\begin{center}
\begin{tabular}{rrrr}
\hline
Dataset & Number of proteins & Length (residues) & Source\\
\hline
small & 4 & 56 & \cite{alexander2009minimal,he2012mutational}\\
medium & 100 & 77$\sim$734 (average: 296) & \cite{zhao2015thermococcus}\\
large & 19,704 & 50 & \cite{rocklin2017global}\\
\hline
\end{tabular}
\end{center}
\end{table}

\begin{figure}[h]
  \centering
  \includegraphics[width=0.7\textwidth]{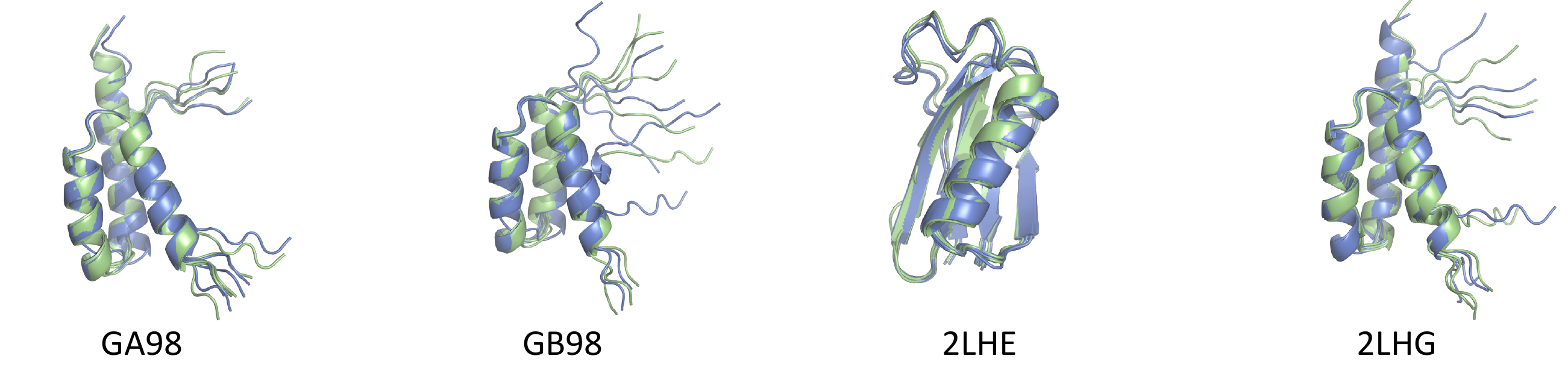}
  \caption{Correctness evaluation with the small dataset. Structure visualised with the ground structures predicted with AlphaFold (blue) and predicted structures with ParaFold (green).}
  \label{fig:Correct}
\end{figure}

\subsection{Evaluation Methods}

We compared \parafold{} with AlphaFold in performance on three datasets of different sizes and lengths. For high throughput protein structure predictions, time, especially the GPU runtime, is the dominant performance metrics of \parafold{} and AlphaFold. We extracted the CPU and GPU runtime recorded in the $timings.json$ file generated by \parafold{} and AlphaFold.


\section{Results and Discussions}

To evaluate the accuracy and efficiency of \parafold{}, we performed comparison across a range of dataset sizes and lengths. We started from the small dataset to show the accuracy and effects of the CPU and GPU optimizations of \parafold{}. We then illustrate the application of \parafold{} for the medium and large dataset to show the efficiency of \parafold{}.

\subsection{Correctness Evaluation}

As for the accuracy, all the optimizations in \parafold{} were implemented without compromising any of the quality of the results. \parafold{} shares the same degree of accuracy as AlphaFold. We avoided any modification of the functional part of AlphaFold. The MSA searches, the databases, the model inference, the three recycling procedures, \etal, were all without any change.

Because of the random seeds used in AlphaFold, and also because of processes like GPU inference that are nondeterministic, the prediction results in AlphaFold (and in \parafold{} as well) may have run variance \cite{alphafoldgithub}.

As shown in Fig \ref{fig:Correct}, the structures of the four proteins in the small dataset predicted by both AlphaFold and \parafold{} were highly identical.

\subsection{CPU acceleration}

\begin{figure}[h]
  \centering
  \includegraphics[width=0.7\textwidth]{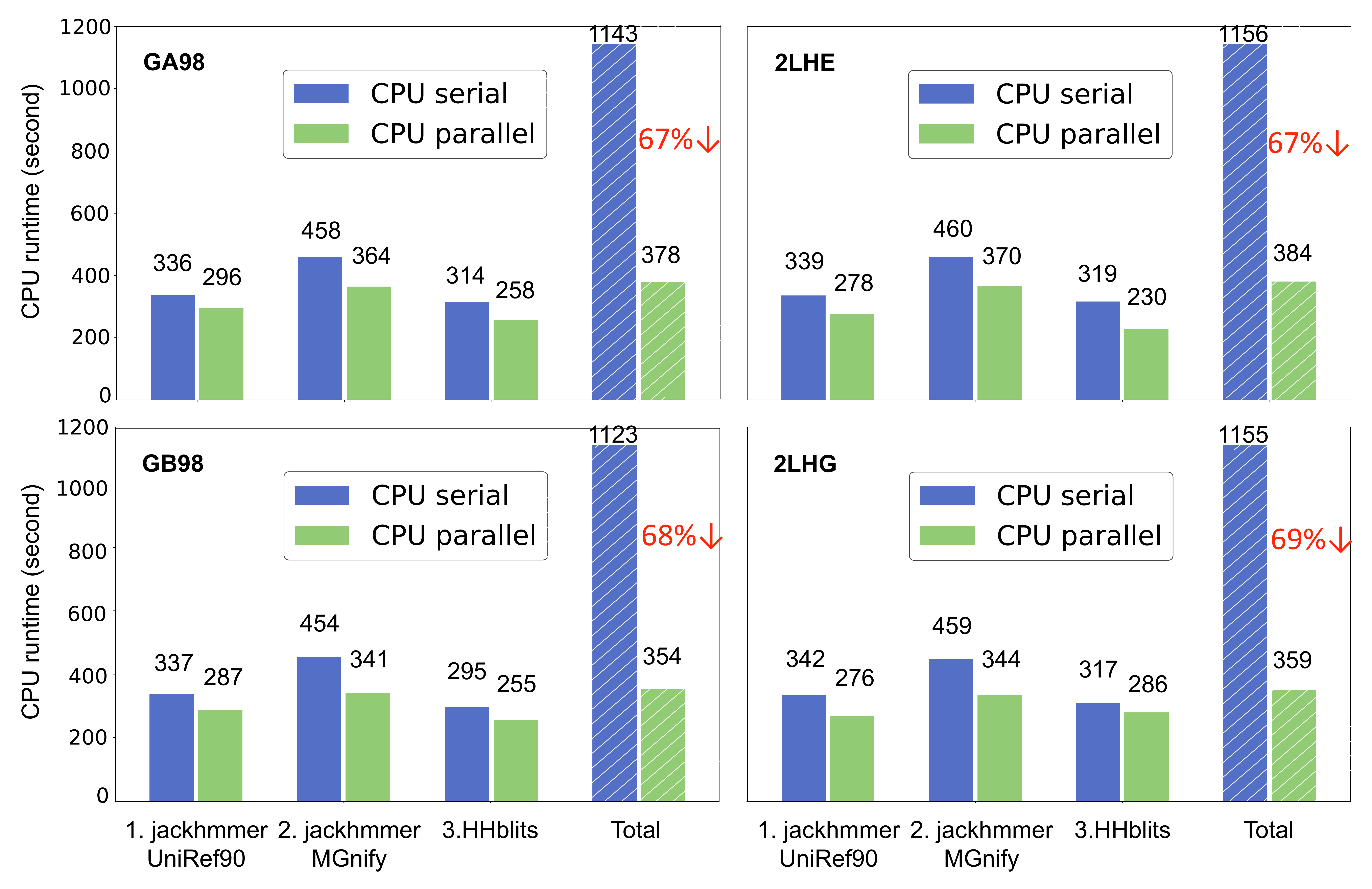}
  \caption{CPU acceleration. For proteins in the small dataset, with parallelism approach in ParaFold, the total CPU runtime was nearly 1/3 of that of the serial process in AlphaFold.}
  \label{fig:cpu}
\end{figure}

With parallel optimization, \parafold{} attained a 3X average speedup on CPUs. Fig \ref{fig:cpu} illustrates the effect of parallel optimization on CPUs with the small dataset. \parafold{} implemented the three MSA searches (1.jackhmmer on UniRef90, 2.jackhmmer on MGnify, 3.HHblits) in parallel on 20 CPU cores. The total CPU runtime achieved $\sim$68\% reduction in \parafold{}, comparing with the original serial process.

\begin{figure}[h]
  \centering
  \includegraphics[width=0.8\textwidth]{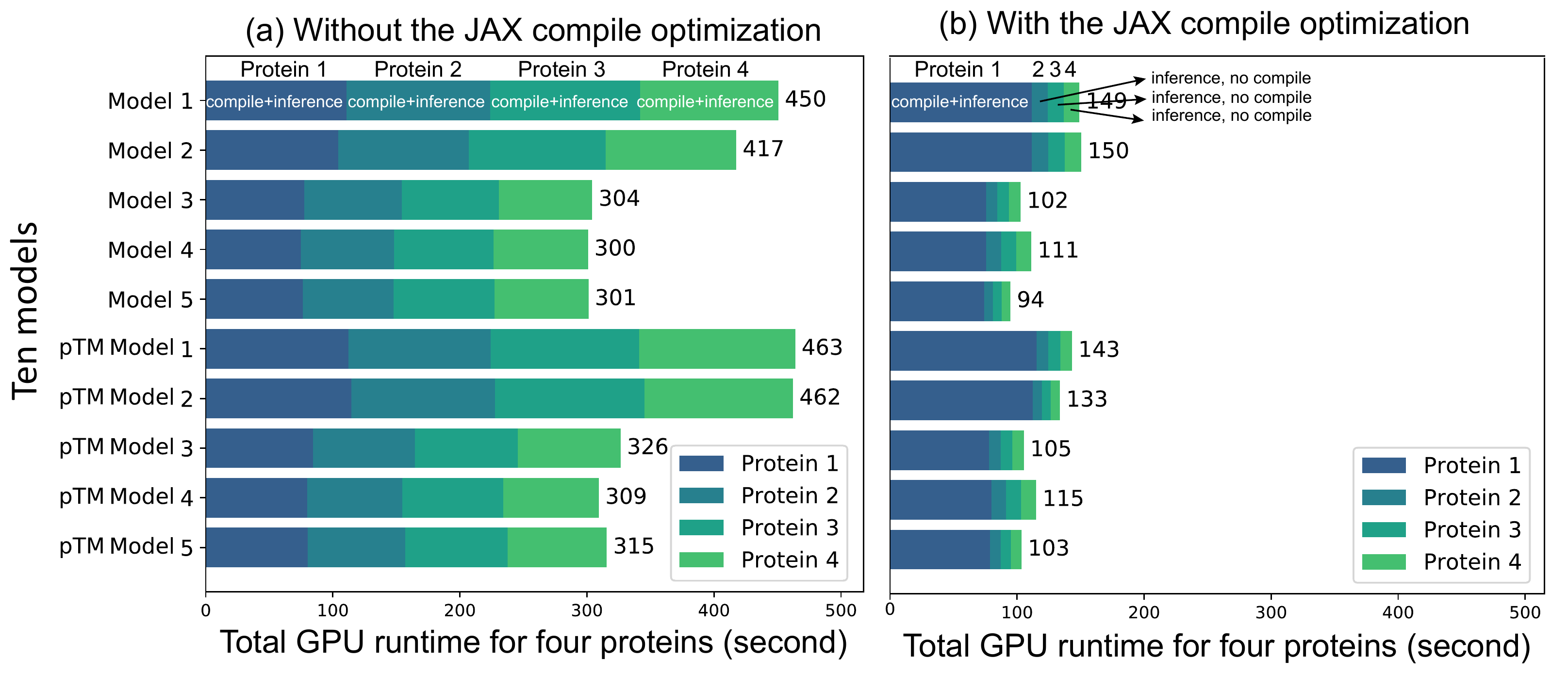}
  \caption{GPU acceleration. Total GPU runtime for four proteins in batch inferences on one V100 GPU. The average time for each compilation was $\sim$60 seconds. (a) Without the JAX compile optimization, JAX compiled 40 times. (b) With the JAX compile optimization, only the first protein's ten models were compiled, thus JAX compiled only 10 times.}
  \label{fig:jax}
\end{figure}

\subsection{GPU acceleration}

\begin{figure}[h]
  \centering
  \includegraphics[width=0.7\textwidth]{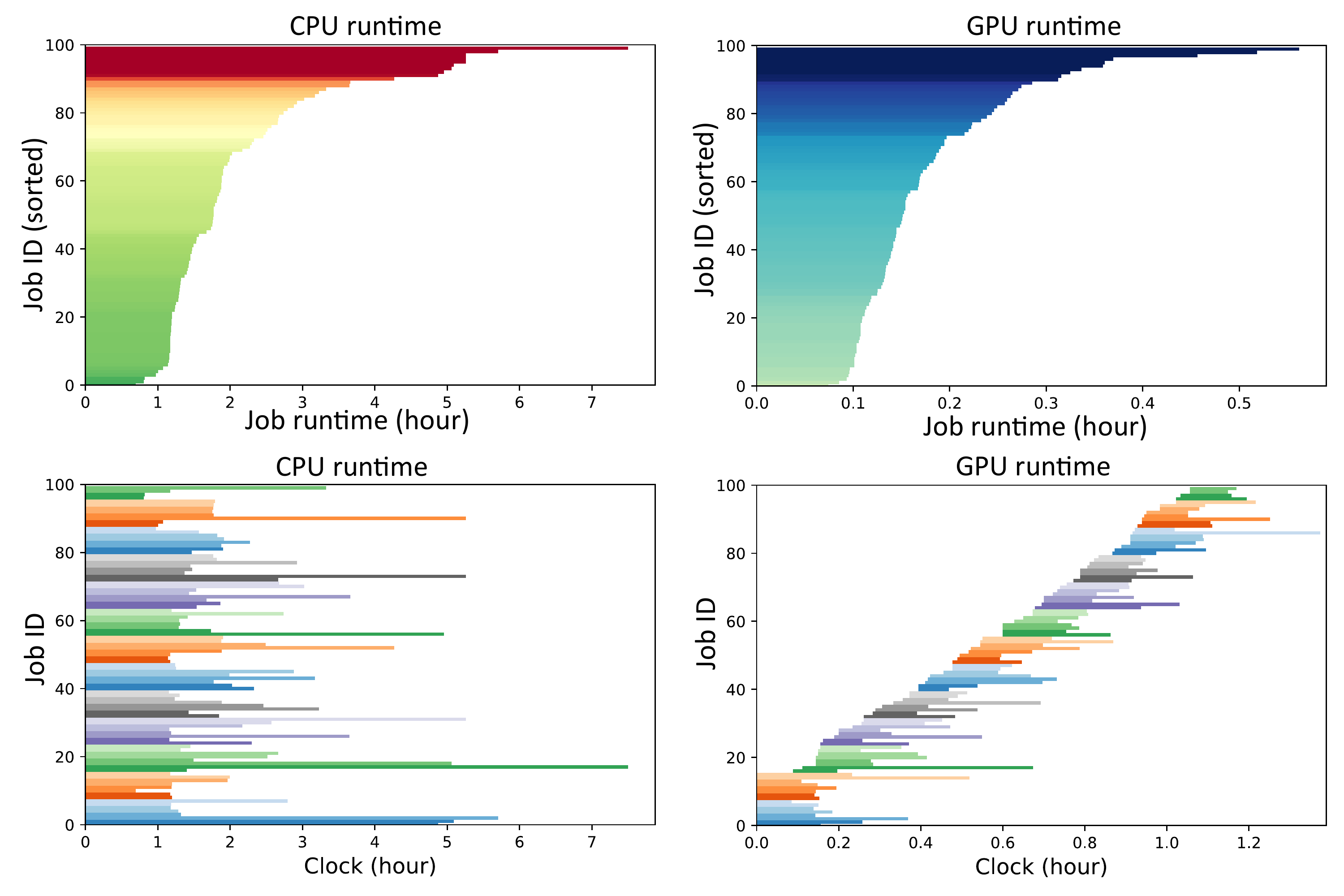}
  \caption{CPU and GPU runtime of ParaFold predictions of the medium dataset. (a) The CPU runtime for each protein ranged from 1 to 7 hours, with an average of $\sim$2 hours. 8 CPU cores were assigned to each job to process one protein's MSA construction. (b) The GPU runtime ranged from 0.1 hour to 0.6 hour, with an average of $\sim$0.2 hour. One DGX-2 was assigned for these 100 proteins' model 1 inference.}
  \label{fig:medium}
\end{figure}

With optimization on GPUs to avoid JAX recompilation, \parafold{} largely reduced GPU runtime for proteins with similar lengths. The speedup was attained by avoiding JAX recompilation in batch inferences. Fig \ref{fig:jax} shows the GPU runtime for four proteins with similar lengths in the small dataset. To predict with 10 models (model 1-5, pTM model 1-5), JAX compiled 10 times for the four proteins in \parafold{}, compared with 40 times in AlphaFold. \parafold{} reused the models compiled for the first protein, without triggering recompilation for the rest proteins in the task.

As shown in Fig \ref{fig:jax}, the average time for JAX compilation was $\sim$60 seconds. Therefore, for each protein (except the first protein), the compile time was reduced by 10$\times$60 seconds. In massive inferences, this optimization resulted in a notable reduction of GPU runtime, as shown later with the large dataset.

\subsection{Pipeline efficiency with the medium dataset}

We show the speed of prediction with \parafold{} across the medium dataset of 100 proteins (Fig \ref{fig:medium}). The lengths of proteins in the medium dataset range from 77 to 734 residues, representing a typical usage scenario with AlphaFold and \parafold{}.

Fig \ref{fig:medium}(a) illustrates the CPU and GPU runtime of \parafold{}. The CPU runtime for each protein ranged from 1 to 7 hours, with an average of $\sim$2 hours, mainly depending on the protein's length. For each CPU job, 8 cores were assigned to process one protein's MSA construction. 

Fig \ref{fig:medium}(b) indicates the GPU runtime for model 1 inference of these proteins on one DGX-2 (GPU 01-16). The GPU runtime ranged from 0.1 hour to 0.6 hour, with an average of $\sim$0.2 hour. For each protein, the model 1 inference was performed on one V100/32G GPU. As shown in the left corner of the bottom figure, 16 jobs were simultaneously allocated for 16 proteins' model 1 inference.

\subsection{Pipeline efficiency with the large dataset}

\begin{figure}[h]
  \centering
  \includegraphics[width=1\textwidth]{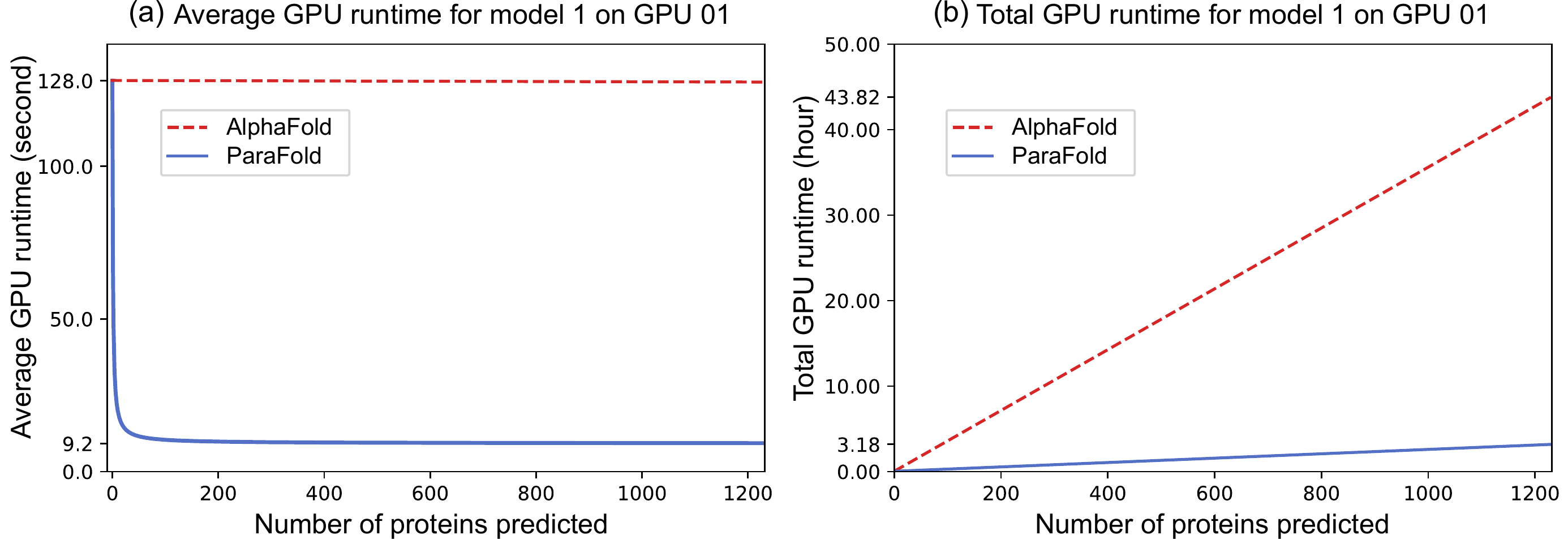}
  \caption{GPU runtime for proteins in the large dataset. We performed the inferences of 19,704 proteins on one DGX-2 (GPU 01-16) using ParaFold (solid line). 1,231 proteins were processed in a single task on GPU 01. The dash line indicates the estimation of GPU runtime if using AlphaFold, based on our calculation result that the average GPU runtime for one protein (50 residues) using AlphaFold for model 1 inference was 128 seconds.}
  \label{fig:GPU_20000}
\end{figure}

\parafold{} accomplished model 1 inference of $\sim$20,000 proteins (50 residues) in 5.4 hours on one NVIDIA DGX-2.  Model 1 inferences of 19,704 proteins in the large dataset were run in 16 tasks, each task for $\sim$1,232 proteins assigned with one V100/32G GPU. The average GPU runtime for one protein was 13.8 seconds (Table \ref{tab:large}).

\parafold{} reduced 99.5\% of the total GPU runtime for these $\sim$20,000 proteins, compared with AlphaFold when taking both MSA generation and model inference into account. To predict these proteins on one NVIDIA DGX-2, AlphaFold would take 1,129 hours. The average GPU runtime of one protein using AlphaFold in our test was 3,312 seconds on one V100 GPU, including 128 seconds for model 1 inference and 3,184 seconds for workload without the need of GPUs.

\begin{table}
\caption{GPU runtime for predictions using model 1 with the large dataset on one NVIDIA DGX-2}
\label{tab:large}
\begin{center}
\begin{tabular}{ccccc}
\hline
Device & Number of Proteins & Total runtime (hour) & Average runtime (second) \\
\hline
GPU 01 & 1,232               & 3.9                & 11.5                   \\
GPU 02 & 1,232               & 5.4                & 15.7                   \\
GPU 03 & 1,232               & 3.9                & 11.5                   \\
GPU 04 & 1,232               & 5.1                & 14.8                   \\
GPU 05 & 1,232               & 3.9                & 11.5                   \\
GPU 06 & 1,232               & 5.0                & 14.7                   \\
GPU 07 & 1,232               & 3.8                & 11.2                   \\
GPU 08 & 1,232               & 5.0                & 14.6                   \\
GPU 09 & 1,232               & 5.3                & 15.5                   \\
GPU 10 & 1,232               & 5.0                & 14.5                   \\
GPU 11 & 1,232               & 3.9                & 11.5                   \\
GPU 12 & 1,232               & 5.0                & 14.7                   \\
GPU 13 & 1,232               & 5.2                & 15.2                   \\
GPU 14 & 1,232               & 5.1                & 14.8                   \\
GPU 15 & 1,232               & 5.1                & 14.8                   \\
GPU 16 & 1,224               & 5.0                & 14.8                   \\  
\midrule
Total on DGX-2  & 19,704              & 5.4                & 13.8                   \\
\hline
\end{tabular}
\end{center}
\end{table}

\begin{figure}[h]
  \centering
  \includegraphics[width=0.4\textwidth]{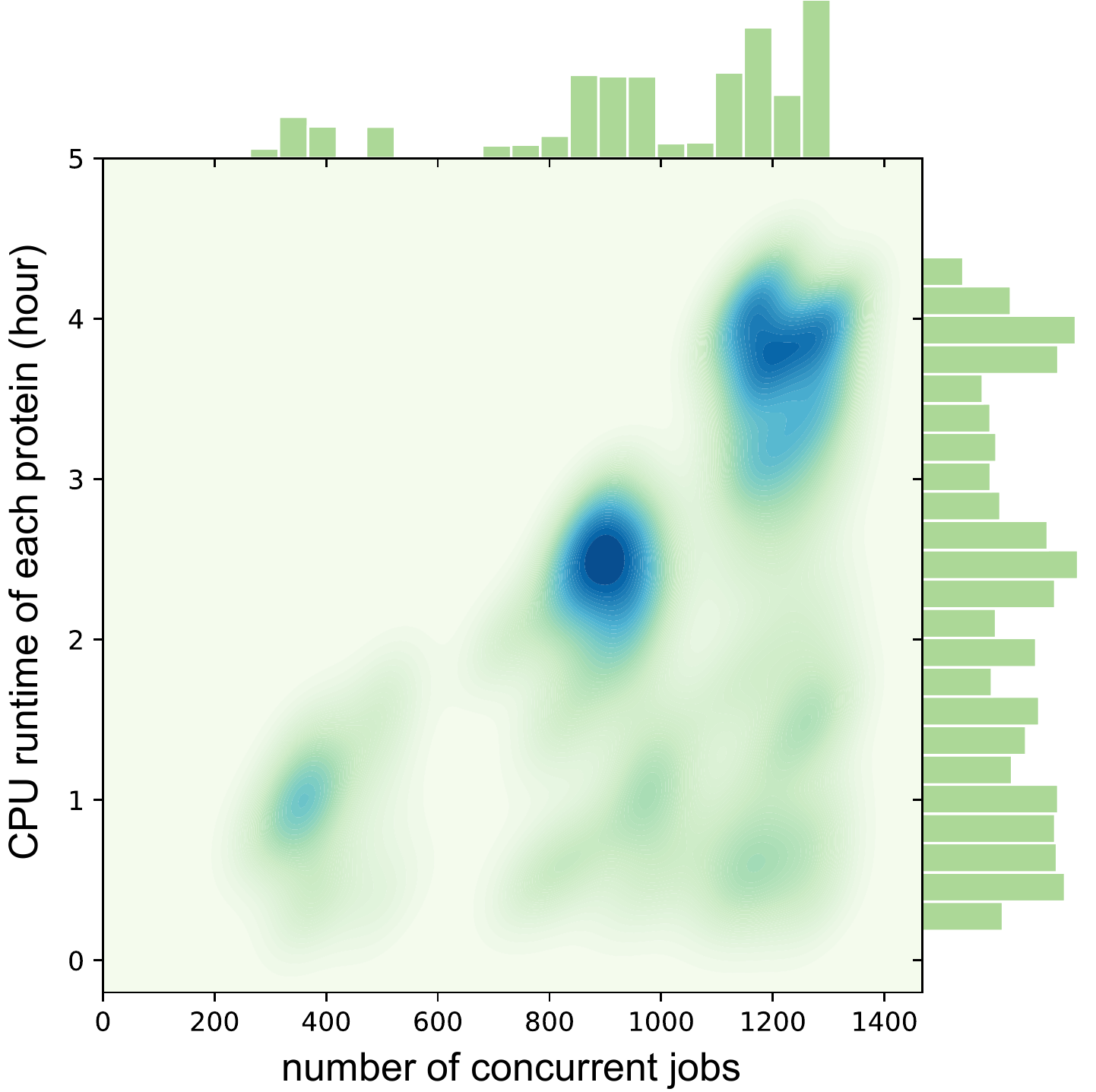}
  \caption{CPU runtime relative to the concurrent jobs for the large dataset. The runtime varies in the range of [0.6, 4.5] hours because of I/O workload on the Lustre file system.}
  \label{fig:io}
\end{figure}

Fig \ref{fig:GPU_20000}(a) shows that, in the GPU stage for model 1 inference, with JAX compile optimization, \parafold{} attained a 13.8X average speedup over AlphaFold. The average GPU runtime of model 1 inference for each protein was 9.2 seconds in \parafold{}, and 128 seconds in AlphaFold. The average GPU runtime decreased as more proteins were processed in the batch inferences, benefiting from the reuse of the models compiled for the first protein.

Fig \ref{fig:GPU_20000}(b) shows that, the total execution time for the 1,231 proteins on GPU 01 was 3.18 hours in \parafold{}, compared with 43.82 hours in AlphaFold. \parafold{} attained a 13.9x total speedup over AlphaFold in model 1 inference.

In the CPU stage, the total resources used for MSA construction amounted to 340,070 CPU core hours. For each protein, the CPU runtime varied with the number of concurrent jobs, ranging from 0.6 hour to 4 hours, as illustrated in Fig \ref{fig:io}. MSA construction took a longer time to process with more concurrent jobs on the Lustre file system of the cluster due to the I/O bottleneck. This is because the genetic search tools such as HHblits in the MSA construction are very I/O intensive. HHblits needs to do many random file access and read operations.

In summary, using \parafold{}, we observed the performance improvement that on the small, medium and large dataset. The GPU runtime was largely reduced, and the high throughput protein structure prediction were effectively empowered by the appropriate usage of the CPU and GPU resources on a supercomputer.


\section{Related work}

Inspired by AlphaFold's recent success, there have now been a few attempts to make AlphaFold faster and more convenient in structure predictions. Contrary to all these work, aimed at providing a parallel version of AlphaFold without any modification of its functions, \parafold{} improves the parallel scalability and efficiency of AlphaFold without any loss in accuracy and virtues in design.

\subsection{ColabFold}

ColabFold \cite{mirdita2021colabfold} combines AlphaFold and RoseTTAFold \cite{baek2021accurate} and uses MMseqs2 \cite{steinegger2017mmseqs2} for a faster MSA generation. ColabFold is an easy-to-use Notebook based environment, and offers many advanced features, such as homo- and hetero-complex modeling and exposes AlphaFold internals. ColabFold's 20-30x faster search and optimized model use allows predicting thousands of proteins per day on a server with one GPU. Coupled with Google Colaboratory, ColabFold becomes a free and accessible platform for protein folding that does not require any installation or expensive hardware.

Though the power of ColabFold is often restricted to the limited GPU resources supplied by Google Colab (16G-GPU to predict a max total length of 1400 residues, and a maximum 12 hours at a time), ColabFold recently offered a solution for use on local servers.

\subsection{End-to-end learning of MSA}

Petti \textit{et al}. \cite{Petti2021.10.23.465204} modified AlphaFold, replacing the MSA with the LAM (Learned alignment module). For a given set of unaligned related sequences, they backpropped through AlphaFold to update the parameters of LAM, maximizing the pLDDT. They demonstrated that by connecting their differentiable alignment module to AlphaFold and maximizing the predicted confidence metric, they can learn MSAs that improve structure predictions over the initial MSAs. This work highlights the potential of differentiable dynamic programming to improve neural network pipelines that rely on an alignment.

\subsection{AlphaDesign}

M Jendrusch \textit{et al}. \cite{jendrusch2021alphadesign} embedded AlphaFold into the design loop as a prediction oracle to enable rapid prediction of completely novel protein monomers starting from random sequences. Their work, AlphaDesign, integrated AlphaFold into target functions to provide high-quality structure prediction and measures of prediction confidence. They then combine this with state-of-the art validation using Rosetta \cite{baek2021accurate} \textit{ab initio} structure prediction and molecular dynamics simulations. 

AlphaDesign modified AlphaFold for single-sequence use by disabling ensembling, templates, extra MSA features and restricting the number of MSA features to the number of monomers modelled. The number of AlphaFold iterations (recycling steps) was kept as a parameter for each optimisation run.


\section{Conclusion and Future Work}

To accelerate scientific discovery in structural biology, \parafold{} offers an efficient and scalable pipeline for predicting protein structures with AlphaFold. \parafold{} builds beyond the initial offerings of AlphaFold by splitting the CPU and GPU parts of the pipeline, providing a speedup for MSA searches with parallel optimization on the CPUs and a speedup for batch inferences by avoiding JAX recompilation on GPUs. 

We evaluated \parafold{} with the three datasets on one NVIDIA DGX-2. The results showed that \parafold{} achieved dramatic reduction in GPU runtime without compromising the quality of the prediction results. High throughput protein structure predictions were effectively empowered by the appropriate usage of the CPU and GPU resources on a supercomputer. \parafold{} completed $\sim$20,000 small proteins structure predictions on one NVIDIA DGX-2 in five hours. \parafold{} is an
open-source software available at GitHub \cite{zurichogithub,hpcgithub}. \parafold{} made rapid and high-quality prediction of protein structures accessible with limited GPU resources, offering effective approach for large-scale structure predictions, leveraging the predictive power of AlphaFold.

Two important tasks are planned in the future, examining to what extend I/O bottlenecks limit the scalability on MSA searches on CPUs, and optimizing JAX for supporting multiple GPUs in model inference.


\section{Acknowledgement}

The computations in this paper were run on the $\pi$ 2.0 supercomputer supported by the Center for High Performance Computing at Shanghai Jiao Tong University. The corresponding author is James Lin (james@sjtu.edu.cn).

\bibliographystyle{ACM-Reference-Format}
\bibliography{acmart}

\end{document}